\documentclass[conference]{IEEEtran}
\ifCLASSINFOpdf
\else
\fi
\usepackage{color}                       
\usepackage[all]{xy}
\usepackage{amsmath,amsthm,amsfonts,amssymb,bm} 
\usepackage{graphicx,psfrag}                    
\usepackage[all]{xy}

\usepackage{float}
\newtheorem{prop}{Proposition}[section]

\newtheorem{lemm}{Lemma}[section]

\begin{document}
%

\title{A class of hopping patterns with minimal collisions}

\author{\IEEEauthorblockN{Qizhi Zhang}
\IEEEauthorblockA{Huawei Technologies Co. Ltd., Beijing, China \\
Email: \{zhangqizhi\}@huawei.com}

}


%


\maketitle

\begin{abstract}
In \cite{VTC} three metrics for hopping pattern performance evaluation is proposed: column period, maximal collision ratio, 
maximal continual collision number, a lower bound of maximal continual collision number is given also. In this paper 
we give a lower bound of maximal collision ratio, a class of hopping pattern whose both maximal collision ratio and
maximal continual collision number fit the lower bounds is constructed also.
\end{abstract}


%
\IEEEpeerreviewmaketitle

\section{Introduction}

In classical wireless communication, devices communicate with each other through base station and core network. 
But the connection capacity of base station is limited. If a lot of devices in a domain have the request of
communication, base station may can not satisfy all the requests.  
D2D ( Device to Device ) communication is a manner that devices communicate with each other through direct
 wireless signal transmitting \cite{SI}.  This technic reduce the load of base station and core network. 
\begin{figure}[H]
\centering
\includegraphics[width=0.45\textwidth]{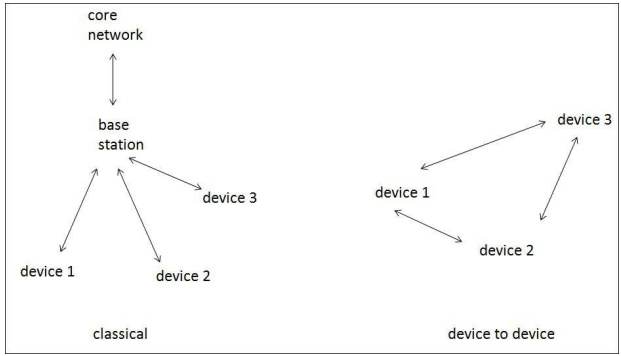}
\caption{Classical wireless communication and D2D communication} 
\end{figure}

OFDMA (Orthogonal Frequency Division Multiple Access) is a  multiple access method that allow multiple users
to use orthogonal subcarrier to transmit signal in same time. This is a suitable multiple access method for D2D
communication, but the half-duplex property need to be considered. It means: a device transmitting own signal
can not receive the signals from other devices in same time.    

 In \cite{QC}, a relative fair method to solve this problem is introduced, the idea is that to let devices "hopping" to 
 avoid always collision. A sequence of periodic frames is set in system. A frame is
divided into $n$ sub frame. In addition, the frequency band of system is split into $m$
 parallel channels using OFDMA, where $m\leq n$. A frequency-time
unit shown in Figure 1 is the basic resource unit for the
device to send or receive a discovery signal. Hence, in one frame, there are $mn$ resource units and each
device use one resource unit to transmit its discovery signal according to the following hopping pattern:
\begin{align*}
& i(t)=i(0) \\
& j(t)=(j(0)+i(0)t) \mod n.
\end{align*}

\begin{figure}[H]
\centering
\includegraphics[width=0.3\textwidth]{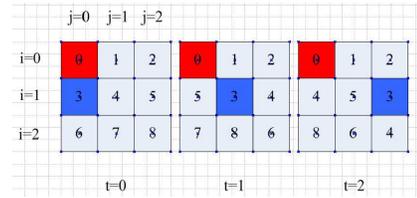}
\caption{Example of hopping pattern}
\end{figure}

The two equations define a UE's transmission frequency-time unit $(i(t), j(t))$ in frame $t$,
 which is decided by the frequency-time unit $(i(0), j(0))$  in frame $0$. For example,
 if UE 0 transmits its discovery signal on red units ,witch $(i(0), j(0))=(0,0), (i(1), j(1))=(0, 0), \cdots$,
and UE 3 transmits its discovery signal on blue units, witch $(i(0), j(0))=(1,0), (i(1), j(1))=(1,1), \cdots$,
though they can not receive each other¡¯s discovery signals in frame $0$, yet in frame $1$ they can receive the signals.

Once hopping pattern is determined, some relation is created between the resource units in differential frames. 
For example, we can see the red frequency-time units as one logical resource, and the blue frequency-time units 
as another logical resource.

In \cite{VTC}, A formalized definition of hopping pattern is given. In which a hopping pattern is defined as a sequence of maps:
\begin{align}
	(i(t),j(t)):S \longrightarrow I \times J  \quad \mbox{ for any } t\in \mathbb{Z}_{\geq 0}	,
\end{align}
where $S$ is the set of logical discovery resources, and contains $|I|\times|J|$ elements, I is the set of frequency-domain  locations, J is the sets of time-domain locations.

\cite{VTC} also give three metrics to evaluate hopping patterns: {\bf column period}, {\bf maximal collision ration},
 {\bf maximal continual collision number} as follows:

Let $\{(i(t), j(t))\}_{t \in \mathbb{Z}_{\geq 0}}$ be a hopping pattern. If there exists a positive integral
 number $T$ satisfying the following condition (P), the hopping pattern is called "be column periodical".

(P): for any $s, s' \in S, t\in \mathbb{Z}$,
\begin{align*}
j(t)(s)=j(t)(s') \mbox{ if and only if } j(t+T)(s)=j(t+T)(s').
\end{align*}

If the hopping pattern is column periodical, the minimal positive
 integral number $T$ satisfying the condition (P) is called the {\bf column period} of the hopping pattern; otherwise the hopping pattern is called has column period infinity.

If for any $s, s' \in S$, there exists a real number $\rho _{s,s'}$ such that
\begin{align*}
\frac{1}{t_b}
\sharp \{t=0, 1, \cdots, t_b-1 \mid  j(t)(s)=j(t)(s') \}
\end{align*}
converges in probability to $\rho _{s,s'}$, the value
\begin{align*}
\begin{array}{c}
\mbox{max} \\
s, s'\in S, s\neq s'
\end{array} \rho_{s,s'}
\end{align*}
is called the {\bf maximal collision ratio} of this pattern.

The {\bf maximal continual collision number} is defined as
\begin{align*}
\mbox{max} \{l=1, 2, ... \mid  &  \mbox{ There exist } s, s', t \mbox{ such that } \\
&j(t)(s)=j(t)(s'), \\
& j(t+1)(s)=j(t+1)(s'), \\
& \cdots, \\
&j(t+l-1)(s)=j(t+l-1)(s')\},
\end{align*}
whose value is either a positive integral number or $\infty$;

A lower bound of the maximal continual collision number is given in \cite{VTC}:

\begin{prop}
\label{continual}
The maximal continual collision number of any hopping pattern on frame structure $\mathbb{Z}/m\mathbb{Z} \times \mathbb{Z}/n\mathbb{Z}$ is greater than or equal to
 $\log _n(m)$. 
\end{prop}

Moreover, some example fitting the  lower bound of maximal continual collision number is given in \cite{VTC}. The hopping pattern proposed in \cite{3GPP_QC}
 fit the lower bound of maximal continual collision number in some situation also. 
 But for maximal collision ratio there is not a result about the lower bound.

In this paper, we give a  lower bound of the maximal collision ratio, and propose a class of pattern that fit the both lower bound of  maximal continual collision number and maximal collision ratio.

\section{The lower bound of maximal collision ratio}
\begin{prop}
The maximal collision ratio of any hopping pattern on frame structure $I \times J$ is greater than or equal to
 $\frac{m-1}{mn-1}$, where $m=|I|, n=|J|$. Moreover, if a pattern fit the lower bound, the number of devices transmitting discovery signal in every time-domain location must be same.
\end{prop}

{\bf Proof.} For any logical discovery resources $s, s'$ and positive integral number $t_b$, let
\begin{align*}
\rho _{s, s'}^{t_b}:=\sharp \{t=0, 1, \cdots, t_b-1 \mid  j(t)(s)=j(t)(s') \}.
\end{align*}
Then we have the maximal collision ratio
\begin{align*}
\max _{s \neq s'} \rho _{s, s'} =\max_{s \neq s'}  \lim _{t_b \rightarrow \infty} \frac{\rho _{s, s'}^{t_b}}{t_b}
\end{align*}
The maximal value should be not less than the average value, hence
\begin{align*}
\max _{s \neq s'} \rho _{s, s'} & \geq \frac{1}{mn(mn-1)} \sum_{s \neq s'}  \lim _{t_b \rightarrow \infty} \frac{\rho _{s, s'}^{t_b}}{t_b} \\
& =\frac{1}{mn(mn-1)} \lim _{t_b \rightarrow \infty} \sum_{s \neq s'} \frac{\rho _{s, s'}^{t_b}}{t_b}.
\end{align*}
Let
\begin{align*}
\rho _{s, s'}(t):=
\left\{
\begin{array}{cl}
1 & \mbox{ if } j(t)(s)=j(t)(s') \\
0 & \mbox{ otherwise }
\end{array}
\right.
\end{align*}
Then we have
\begin{align*}
\sum_{s \neq s'} \rho _{s, s'}^{t_b}=\sum_{s \neq s'} \sum _{t=0} ^{t_b} \rho _{s, s'} (t)=\sum _{t=0} ^{t_b}\sum_{s \neq s'}\rho _{s, s'} (t).
\end{align*}
From the lemma \ref{disjoint} below, we know that
\begin{align}
\label{3423}
\sum_{s \neq s'}\rho _{s, s'} (t) \geq m^2n-mn
\end{align}
for any $t$. Therefore we have
\begin{align*}
\sum_{s \neq s'} \rho _{s, s'}^{t_b} \geq t_b(m^2n-mn),
\end{align*}
and hence
\begin{align*}
\max _{s \neq s'} \rho _{s, s'} \geq \frac{1}{mn(mn-1)} (m^2n-mn)=\frac{n-1}{mn-1}.
\end{align*}
If a pattern has  maximal collision ratio $\max _{s \neq s'} \rho _{s, s'}=\frac{m-1}{mn-1}$, the inequality (\ref{3423}) should be equality for any $t$. Hence the number of devices transmitting discovery signal in every time-domain location must be same by lemma \ref{disjoint}.         \qed

\begin{lemm}
\label{disjoint}
Let $X$ be a set consists of $mn$ elements, and be divided into $n$'s disjoint parts $X=\coprod_ {j=1} ^n X_i$.
Let
\begin{align*}
\rho _{x, x'}=
\left\{
\begin{array}{ll}
1    &  \mbox{ if }x \mbox{ and } x' \mbox{ in same part } \\
0   & \mbox{ otherwise }
\end{array}
\right.
\end{align*}
Then $\sum _{x \neq x' \in X} \rho _{x, x'} \geq m^2n-mn$. The equality sign is fitted if and only if the numbers of elements in every parts $X_j$ are same.
\end{lemm}
{\bf Proof.}
\begin{align*}
&\sum _{x \neq x' \in X} \rho _{x, x'} \\
=&\sum_ {j=1} ^n  \sum _{x \neq x' \in X_j} \rho _{x, x'}  \\
=& \sum_ {j=1} ^n   |X_j|(|X_j|-1)\\
=& \sum_ {j=1} ^n   |X_j|^2-\sum_ {j=1} ^n   |X_j|
\end{align*}
We know that $\sum_ {j=1} ^n   |X_j|=|X|=mn$, and
\begin{align*}
\sqrt {\frac{\sum_ {j=1} ^n   |X_j|^2}{n}} \geq \frac{\sum_ {j=1} ^n   |X_j|}{n}=m,
\end{align*}
which implies
\begin{align*}
&\sum _{x \neq x' \in X} \rho _{x, x'} \geq m^2n-mn.
\end{align*}
 The equality sign is fitted if and only if the cardinality of every $X_j$s  are same. \qed

\section{A class of pattern fitting both lower bounds}

In \cite{VTC}, authors gave a class of pattern whose maximal continual collision number fit to the lower bound  $\log _n(m)$ and whose maximal collision ratio equal to $\frac{1}{n}$ which close to the lower bound $\frac{m-1}{mn-1}$, where $m=|I|, n=|J|$. In this section we propose a class of hopping pattern whose maximal continual collision number fit to the lower bound  $\log _n(m)$ as well as the maximal collision ratio fit to the lower bound $\frac{m-1}{mn-1}$.

Suppose the number $n=q$ be powers of a prime number $p$, $m=q^r$ be a power of $q$. Fix a bijection $I \simeq \mathbb{F}_q^{(r)}$ and
$J \simeq \mathbb{F}_q$. We will always use a column vector in $\mathbb{F}_q^{(r)}$ to represent a frequency domain location, and use  a column vector in $\mathbb{F}_q$ to represent a time domain location in following.

Let and $f(x)=x^{r+1}+a_1x^{r}+ \cdots +a_{r+1}$ be an irreducible polynomial on $\mathbb{F}_q$ satisfying the following condition (G):

(G): The minimal integral number $c$ satisfying
\begin{align*}
x^{c} \equiv 1 \mod f(x) \mbox{     ( as the polynomials in $\mathbb{F}_q[x]$ )}
\end{align*}
is $p^{r+1}-1$. ( In fact, the image of such $x$ under the ring homomorphism  $\mathbb{F}_p[x] \longrightarrow \frac{\mathbb{F}_p[x]}{(f(x))} \backsimeq \mathbb{F}_{p^{r+1}}$ is a generator of the multiplicative group $\mathbb{F}_{p^{r+1}}^\times$. )
\qed

Let $(i(0),j(0)): S \longrightarrow \mathbb{F}_q^{(r)} \times \mathbb{F}_q$ be any bijection,
and define $\{(i(t), j(t))\}_{t \in \mathbb{Z}_{\geq 0}}$ as follows:

\begin{align*}
\left(
\begin{array}{c}
i(t+1) \\
j(t+1)
\end{array}
\right)=
A
\left(
\begin{array}{c}
i(t) \\
j(t)
\end{array}
\right)
\end{align*}
where $A$ is the companion  matrix of $f(x)$ on $\mathbb{F}_q$, i.e
\begin{align*}
A=
\left(
\begin{array}{ccccc}
0 & 0 & \cdots & 0 & -a_{r+1} \\
1 & 0 & & \vdots & -a_{r} \\
0 & 1 & \ddots & \vdots \\
\vdots & \ddots & \ddots & 0 & \vdots \\
0 & \cdots & 0 & 1 & -a_1
\end{array}
\right)
\end{align*}

\begin{prop}
The hopping pattern $\{(i(t), j(t))\}_{t \in \mathbb{Z}_{\geq 0}}$ defined previously has column period $\frac{mn-1}{n-1}$,
 maximal continual collision number $\log _n (m)$ and maximal collision ratio $\frac{m-1}{mn-1}$.
\end{prop}

{\bf Proof.}

{\bf Column period.}
It is easy to see that
\begin{align*}
\left(
\begin{array}{c}
i(s+t) \\
j(s+t)
\end{array}
\right)=
A^t
\left(
\begin{array}{c}
i(s) \\
j(s)
\end{array}
\right)
\end{align*}
for any integral number $s,t$.

By lemma \ref{4567}.{\bf a} below we know that
\begin{align*}
\left(
\begin{array}{c}
i(s+\frac{q^{r+1}-1}{q-1}) \\
j(s+\frac{q^{r+1}-1}{q-1})
\end{array}
\right)=
\left(
\begin{array}{c}
yi(s) \\
yj(s)
\end{array}
\right)
\end{align*}
for some $y \in \mathbb{F}_q^\times$. Hence the column period is a divisor of $\frac{q^{r+1}-1}{q-1}$.

Suppose the column period is $T$, it means that $A^T$ has the form
\begin{align*}
A^T=
\left(
\begin{array}{cc}
\tilde{A} & \beta \\
0 & \lambda\\
\end{array}
\right)
\end{align*}
where $\tilde{A}$ is a square matrix of degree $r$ on $\mathbb{F}_q$, $\beta$ is a column vector of dimension $r$ on $\mathbb{F}_q$,
 and $\lambda$ is a element in $\mathbb{F}_q^\times$.

Let $Y=\{(0, 0, \cdots, 0, y_{r+1}) \in \mathbb{F}_q^{r+1} \mid y_{r+1} \in \mathbb{F}_q^\times\}$. It is easy to see that
\begin{align*}
YA^T \subset Y.
\end{align*}
Hence we know that
\begin{align*}
A^T \in \left\{ yI_{r+1} \mid y\in \mathbb{F}_q^\times \right\}
\end{align*}
by lemma \ref{4567}.{\bf d} below. Hence $T$ is divided by $\frac{q^{r+1}-1}{q-1}$.

Hence $T=\frac{q^{r+1}-1}{q-1}=\frac{mn-1}{n-1}$.

{\bf Maximal collision ratio.}
We know that the column period of the hopping pattern is a divisor of $q^{r+1}-1$. For any two different logical discovery resources
$s$ and $s'$, we have
\begin{align*}
\left(
\begin{array}{c}
i(0)(s)-i(0)(s') \\
j(0)(s)-j(0)(s')
\end{array}
\right) \neq 0
\end{align*}
and
\begin{align*}
\left(
\begin{array}{c}
i(t)(s)-i(t)(s') \\
j(t)(s)-j(t)(s')
\end{array}
\right)=
A^t
\left(
\begin{array}{c}
i(0)(s)-i(0)(s') \\
j(0)(s)-j(0)(s')
\end{array}
\right) .
\end{align*}
The lemma \ref{4567}.{\bf b} below shows that
\begin{align*}
A^t
\left(
\begin{array}{c}
i(0)(s)-i(0)(s') \\
j(0)(s)-j(0)(s')
\end{array}
\right) \quad \mbox{ for } t=0, 1, \cdots, q^{r+1}-2
\end{align*}
take every values in $\mathbb{F}_q^{(r+1)} \setminus \{0\}$ once. On the other hand, there are $q^{r}-1$'s elements in
$\mathbb{F}_q ^{(r+1)} \setminus \{0\} $ with last coordinate $0$, which means the collision number of the two logical discovery resources $s$ and $s'$ is $q^r-1$ in continual $q^{r+1}-1$'s frames. Hence the maximal collision ratio of this pattern is $\frac{q^r-1}{q^{(r+1)}-1}=\frac{m-1}{mn-1}$.

{\bf Maximal continual collision number.} We know the maximal continual collision number is greater than or equal to $\log_n (m)=r$ by proposition \ref{continual}.
Suppose it is greater than $r$, then there are two different logical resources $s$ and $s¡ä$,
and an integral number $t$ such that
\begin{align*}
j(t)(s)&=j(t)(s') \\
j(t+1)(s)&=j(t+1)(s') \\
&\vdots \\
j(t+r)(s)&=j(t+r)(s')
\end{align*}
Let $\beta:=i(t)(s)-i(t)(s')$, then we have
\begin{align*}
(0, 0, \cdots, 1)A^k
\left(
\begin{array}{c}
\beta \\
0
\end{array}
\right)=0
\end{align*}
for $k=0, 1, \cdots, r$. Let $R$ be a matrix of degree $r+1$ whose $k'th$ row is $(0, 0, \cdots, 1)A^k$, then we have
\begin{align*}
R
\left(
\begin{array}{c}
\beta \\
0
\end{array}
\right)=0
\end{align*}
But we know the rows of R are linear independent on $\mathbb{F}_q$ from lemma \ref{4567}.{\bf e} below,
 hence $R$ is a non-singular matrix, which implies $\beta=0$. It is contradictory to that $s$ and $s'$ 
 are different logical resources.
 \qed

\begin{lemm}
\label{4567}
Suppose $A$ is defined as before, then we have:

{\bf a.} The multiplication group $<A>$ generate by $A$ is a cyclic group of degree $q^{r+1}-1$; and the
multiplication group $<A^{\frac{q^{r+1}-1}{q-1}}>$ generate by $A^{\frac{q^{r+1}-1}{q-1}}$ equal to $\left\{ yI_{r+1} \mid y\in \mathbb{F}_q^\times
\right\}$.

{\bf b.} $<A>$  faithfully and transitively acts on $\mathbb{F}_q^{(r+1)} \setminus \{0\}$ buy left multiplication. ( A group left action $G \times X \longrightarrow X$ is called faithful and transitive if and only if for any $x\in X$, the map $g \mapsto gx$ is a bijection from $G$ to $X$.)

{\bf c.} $<A>$  faithfully transitively acts on $\mathbb{F}_q^{r+1} \setminus \{0\}$ buy right multiplication.
( A group right action $X \times G \longrightarrow X$ is called faithful and transitive if and only if for any $x\in X$, the map $g \mapsto xg$ is a bijection from $G$ to $X$.)

{\bf d.} Let $Y=\{(0, 0, \cdots, 0, y_{r+1}) \in \mathbb{F}_q^{r+1} \mid y_{r+1} \in \mathbb{F}_q^\times\}$, and
$H_Y:=\{ \sigma \in <A> | Y\sigma \subset Y\}$, then we have
\begin{align*}
H_Y=\left\{ yI_{r+1} \mid y\in \mathbb{F}_q^\times
\right\}
\end{align*}
where $I_{r+1}$ is the identity matrix of degree $r+1$.

{\bf e.} For any $\alpha \in \mathbb{F}_q^{r+1} \setminus \{0\}$, the row vectors $\alpha A^k, k=0, 1, \cdots, r,$ are linear 
independent on $\mathbb{F}_q$. 
\end{lemm}

{\bf Proof.}

{\bf a.} Because under the isomorphism $\mathbb{F}_q[A] \simeq \mathbb{F}_q[x]/(f(x)) \simeq \mathbb{F}_{q^{r+1}}$, the image
$\bar{x}$ of $A$ is a generator of the multiplicative group $\mathbb{F}_{q^{r+1}} ^\times$, we know
$A$ has multiplicative order $q^{r+1}-1$. On the other hand, $\bar{x}^{\frac{q^{r+1}-1}{q-1}}$ generate
the multiplicative subgroup $\mathbb{F}_q ^\times$, which is corresponding to $\left\{ yI_{r+1} \mid y\in \mathbb{F}_q^\times
\right\}$ in $\mathbb{F}_q[A]$.

{\bf b.}
Consider the commutative diagram
\begin{align*}
\xymatrix{\ar @{} [dr] |{}
<A> \ar[d]^\wr  & \times  & \mathbb{F}_q^{(r+1)}  \setminus \{0\} \ar[d]^\wr \ar[r] & \mathbb{F}_q^{(r+1)}  \setminus \{0\} \ar[d]^\wr \\
\mathbb{F}_{q^{r+1}}^\times & \times & \mathbb{F}_{q^{r+1}}^\times \ar[r] & \mathbb{F}_{q^{r+1}}^\times }
\end{align*}
The left vertical map defined by $A \mapsto \bar{x}$ is a isomorphism between groups. The middle vertical map defined by
\begin{align*}
\left(
\begin{array}{c}
a_0 \\
a_1 \\
\vdots \\
a_r
\end{array}
\right)  \mapsto
(1, \bar{x}, \cdots, \bar{x}^r)
\left(
\begin{array}{c}
a_0 \\
a_1 \\
\vdots \\
a_r
\end{array}
\right)
\end{align*}
is a bijection between sets. The right vertical map is same as the middle one.

The map in up line is defined by the multiplication of matrix, the map in down line is defined by  the multiplication in finite fields.

Because $\mathbb{F}_{q^{r+1}}^\times$ acts on $\mathbb{F}_{q^{r+1}}^\times$ faithfully and transitively, we have $<A>$  faithfully transitively acts on $\mathbb{F}_q^{(r+1)} \setminus \{0\}$.

{\bf c.} It is similar to the proof of {\bf b}.

{\bf d.} Let $X=\mathbb{F}_q^{(r+1)} \setminus \{0\}$, then we have
\begin{align*}
X \subset \bigcup _{\sigma \in <A>} Y\sigma = \bigcup _{\sigma \in <A>/H} Y\sigma
\end{align*}
Therefore
\begin{align*}
|X| \leq |Y| \times |<A>/H|
\end{align*}
But we know that $|X|=|A|$, hence
\begin{align*}
|H|=|Y|=q-1.
\end{align*}
But it easy to see that $\left\{ yI_{r+1} \mid y\in \mathbb{F}_q^\times
\right\} \subset H_Y$, hence we know
\begin{align*}
H_Y=\left\{ yI_{r+1} \mid y\in \mathbb{F}_q^\times
\right\}.
\end{align*}

{\bf e.} Suppose there are $a_0, a_1, \cdots, a_n \in \mathbb{F}_q$ not all zero such that 
\begin{align*}
\sum _{k=0} ^r a_k \alpha A^k =0,
\end{align*}
we will show that there is a contradiction.

Let $g(x)=a_0+a_1x+a_2x^2+ \cdots +a_r x^r$, then we have
\begin{align*}
\alpha g(A) =0.
\end{align*} 
On the other hand, we have
\begin{align*}
\alpha f(A) =0
\end{align*} 
because $A$ is the companion matrix of $f(x)$. Hence we have
\begin{align*}
\alpha h(A)=0
\end{align*}
where $h(x)$ is the greatest common divisor of $f(x)$ and $g(x)$. Because $f(x)$ is irreducible and $g(x)\neq 0$
 and $\deg g(x) \leq r$, we know that $f$ and $g$ are co-prime. Hence we have $h(x)=1$, and then
\begin{align*}
\alpha =0,
\end{align*}
which is contradictory to $\alpha \in \mathbb{F}_q^{r+1} \setminus \{0\}$.
\qed

\section{Conclusion and future work}
In this paper, we gave a lower bound of the maximal collision ratio of hopping patterns.
Then a class of hopping pattern is constructed. Both maximal collision ratio and maximal continual collision number of these
pattern fit to the lower bounds. These pattern apply to the frame structure that the number $n$ of time-domain resources is
a power of a prime number, and the number $m$ of frequency-domain resources is a power of $n$.

It is still an open problem that how to construct a hopping pattern fitting the both lower bounds 
of maximal collision ratio and maximal continual collision number for more general 
frame structure.






%

\end{document}